
%


\input phyzzx


\def\j{\journal}

\def\a{\alpha_s}
\def\mt{m_t}
\def\gf{G_F}

\def\4pi{(4 \pi)}
\def\dslash{\hbox{$\partial$\kern-0.5em\raise0.3ex\hbox{/}}}
\def\Dslash{\hbox{$D$\kern-0.5em\raise0.3ex\hbox{/}}}
\def\Aslash{\hbox{$A$\kern-0.5em\raise0.3ex\hbox{/}}}
\def\Gslash{\hbox{$G$\kern-0.5em\raise0.3ex\hbox{/}}}
\def\m{\mu}
\date{CERN-TH.7446/94}
\pubtype{}
\titlepage
\title{ {QCD corrections to large-$m_t$ electroweak
effects in $\Delta r$.\break
An  effective  field  theory point of view.}\ \ \foot{Work partially supported
by research
project CICYT-AEN93-0474.}}
\author{S. Peris\foot{On leave from Grup de Fisica Teorica and IFAE,
Universitat Autonoma de Barcelona, Barcelona, Spain. e-mail: PERIS @
CERNVM.CERN.CH}}
\address{Theory Division, CERN,\break CH-1211 Geneva 23\break
Switzerland}
\abstract

By using effective field theory techniques for the standard model, we
discuss the issue of what $\mu$ scale is the appropriate one in the QCD
corrections to the large-$\mt$ electroweak contributions to $\Delta r$.
This needs the
construction of an effective field theory below the top quark. We argue that,
while matching corrections do verify the simple prescription of taking
$\mu \simeq \mt$ in $\a(\m)$, logarithmic (i.e. $\sim \log \mt$) corrections
do not, and require
the use of the running $\a(\mu)$ in the corresponding renormalization group
equation.

\vskip .5in

CERN-TH.7446/94


September 1994

\endpage

The high precision  experiments done at LEP\Ref\Lep{See for instance, R.
Miquel, CERN-PPE/94-70. Talk given at the 22nd Symposium on "Physics with
High Energy Colliders", Tokyo, Japan, March 1994.}
and the SLC\Ref\Slc{K. Abe et al., SLD Collaboration, SLAC-PUB-6456, March
1994.}
 have
recently motivated the interest in the $\a$ corrections to the
large-$m_t$ contributions to observables in the standard model (SM).
We would like to concentrate here on $\Delta r$ for its simplicity.

\Ref\Djouadi{A. Djouadi and C. Verzegnassi\j Phys. Lett. &B195 (87) 265; A.
Djouadi\j Nuovo Cimento &100A (88) 357.}
\Ref\Sergio{See for instance S. Fanchiotti, B. Kniehl and A. Sirlin\j
Phys. Rev. &D48 (93) 307; F. Halzen, B. Kniehl and M.L. Stong, MAD/PH/643,
Lectures presented by F. Halzen at the VI J. Swieca Summer School, Sao
Paulo, Brazil, 1991.}
\Ref\Jegerlehner{F. Jegerlehner, Lectures delivered at
TASI-90, Boulder, Colorado, June 1990.}
\Ref\Yellow{"Physics at LEP 1", yellow report,
CERN 89-08, edited by G. Altarelli, R. Kleiss and C.
Verzegnassi.}
According to refs. [\Djouadi-\Yellow], $\Delta r$ can be expressed as
\foot{There is a typographical error in the $\log M^2_W/m_t^2$
term of $\Delta r(top)$ in ref. [\Yellow], which appears with an overall minus
sign with respect to our expression. We thank F. Jegerlehner for confirming
this.}
$$\eqalign{\Delta r(top)\approx& -{c^2\over s^2}\ {3 \mt^2 \gf \sqrt 2\over
\4pi^2}
\left(1-{\a(\mu)\over \pi}{6+2 \pi^2\over 9}\right)+\cr
&+{g^2\over \4pi^2}\ {1\over 2} \left({c^2\over s^2}-{1\over 3}\right)
\log \left({M_W^2\over \mt^2}\right)
\left(1+ {\a(\mu)\over \pi}\right)\qquad ,\cr}\eqn\A$$
where we have only kept the leading and next-to-leading $\mt$ dependence.

Along with these results, there has appeared the discussion of the $\mu$
scale at which one
is supposed to evaluate $\a(\mu)$ in these expressions\foot{In principle
three scales appear in these loops: $\mt,M_Z,m_b$.}, and
the parameters in terms of which
one ought to express the result, i.e. whether $\overline {MS}$, or
on-shell, etc. For this, a prescription has been designed that
says\refmark{\Sergio} that corrections coming from the $(t,b)$ doublet
should be
computed with $\a(\mt)$. This prescription would then say that in all of the
above
expressions $\a(\mu)$ should be taken as $\a(\mt)$. We would like to explain
what an effective field theory (EFT) point of view shows about this
issue.
We shall see that while
this prescription works for the power-like terms (those that go like $\mt^2$),
the renormalization group (RG) supplies us with a different result for the
logarithmic terms (those that go like $\log \mt$)
\foot{From an effective field theory point of view these
two types of contributions are totally different. While the former (i.e.
power-like) come
from "matching", the latter (i.e. logarithmic) come from "running".
See below.}.

The EFT\Ref\Eft{See, for instance, H. Georgi\j Annu. Rev. Nucl. Part.
Sci. &43 (93) 209 and references therein.}
is a very suitable framework for gaining control over the problem of
the interrelation of scales and quantum corrections just because of the very
nature of an EFT. Indeed, in the process of constructing it, one must
clearly identify the relevant scales appearing in the problem. The
construction goes as follows\Ref\Book{For a recent, very nice and detailed
discussion, see M. Bilenky and A. Santamaria\j Nucl. Phys. &B420 (94) 47.
See also H. Georgi, "Weak Interactions and
Modern Particle Theory", The Benjamin/Cummings Pub. Co., 1984.}
\refmark{\Eft}. For a certain physical process occurring at an
energy scale $E_0$, one starts with the bare Lagrangian (which is valid at
an energy $E>>E_0$, where $E$ is bigger than any particle's mass) and
integrates out all particles whose masses are larger than $E_0$. In this
process, every time one integrates out a particle one does a large mass
expansion, so that
there appears a tower of
higher dimensional operators weighted with the appropriate inverse powers of
the particle's mass one has just integrated out. Consequently,
one has to include new
counterterms to
compensate for the different high energy behavior of the effective theory under
construction with respect to the original full theory, so that both the
effective and the full theory yield the same physics. This is typically
done at a scale $\mu$\foot{For convenience, we shall always work in the
${\overline {MS}}$ scheme. Accordingly, couplings and masses will depend on
the scale $\mu$. Unless otherwise stated,
only the pole mass will have no explicit dependence on
$\mu$ in the text.} equal to the particle's mass to minimize the effect of
large logarithms. Power corrections appear right at this point. This is
called a "matching condition". Below the particle's mass one can then use
the powerful machinery of the renormalization group to obtain the effective
Lagrangian relevant at lower scales; in other words, one "runs" the
Lagrangian down to the lower scales. It is then that one obtains the
typical logarithms between two scales. If in the process of doing so one
encounters another particle, then one again has to integrate it out and find
the
matching condition  at threshold for the corresponding counterterm, so that
the procedure can be iterated until the energy
scale $E_0$ is reached. Only
particles lighter than the scale $\mu$ one is at, at every moment, are to be
included in the running, i.e. in the RG equations. Therefore, in an EFT all
the contributions are either "matching" or "running", and there is nothing
else.

In this paper we shall be concerned with the QCD corrections to the
large-$m_t$ one-loop electroweak corrections. Therefore, for all practical
purposes, one may think as if the top quark were the heaviest particle in
the SM, much heavier than the Higgs boson, which is taken to be
nearly degenerate with the W and
Z. This automatically kills the $\log M_H/M_W$ contributions and leaves the
$\mt^2$ and the $\log \mt/M_W$ ones, which are those we are interested in.

The general philosophy will parallel that so successfully used in the
context of grand unified theories\Ref\Hall{L. Hall\j Nucl. Phys. &B178 (81)
75.}. There is of course a very important  difference, namely that, upon
integration of the top quark, the resulting effective theory will no longer
exhibit an explicit linear SU$_2\times$U$_1$ invariance. This would make a full
account of the corresponding RGE's very cumbersome. Luckily we may
keep only those contributions that are strictly relevant.

Let us start with the full SM at $\mu>\mt$. At $\mu=\mt$, one
integrates the top out obtaining\foot{For instance, in the context of grand
unification, $g$ in the following equation would be the value of the
SU$_2^{EW}$ gauge coupling constant at the scale $\mu=\mt$ as obtained from
the running of the corresponding $\beta$ function between M$_{GUT}$ and
$\mt$.}
$$\eqalign{{\cal L}_{eff}=&\left(1+ g^2 \delta Z_+(\mu)\right) \  W^+_{\mu}
\partial^2 {W^-}^{\mu}+ {g^2\over 4} \left(v^2+\delta v_+^2(\mu)\right)
{W_{\mu}}^+ {{W^-}^{\mu}} +\cr
&+\ {1\over 2}\  \left(1+g^2 \delta Z_3(\mu)+g^2\delta Z_{3Y}(\mu)\right)
\ W^3_{\mu}\partial^2 W_3^{\mu}+\cr
&+\ {1\over 2}\ \left(1+g'^2 \delta Z_Y(\mu)+g'^2\delta Z_{3Y}(\mu)\right)
\ B_{\mu}\partial^2 B^{\mu}+\cr
&+\ {1\over 2}\ \left(gW_3^{\mu}-g'B^{\mu}\right)\
\left[{1\over 4}\left(v^2+\delta
v^2_3(\mu)\right)-\delta Z_{3Y}(\mu)\ \partial^2\right]\
\left(gW^3_{\mu}-g'B_{\mu}\right)+ \cr
&+\ \bar \psi \ i\Dslash(gW^+,gW_3,g'B)\ \psi \ \cr}\eqn\one$$
from the diagrams of fig. 1. Here $\psi$ stands for all the fermions but the
top. Notice that we have dealt with $W_3-B$ mixing by including a
$\partial^2$ operator in the form of a generalized mass term in eq. \one.
This will make the subsequent diagonalization very simple since the neutral
mass eigenstate is still of the form $gW_3-g'B$, like at tree level.
Certainly, there will also be a tower of
higher dimensional operators suppressed by the corresponding inverse
powers of the
top quark mass, but we shall neglect them throughout. Possible four-fermion
operators are irrelevant to the discussion that follows and are also
disregarded.
We also postpone the study of the $Zb\bar b$ vertex to a future
analysis\Ref\future{S. Peris, in preparation.}.

We can now redefine our fields in order to have standard kinetic terms. This
yields
$$\eqalign{&{\cal L}= W_\mu^+ \partial^2 {W^-}^\mu+
{g_+^2(\m)\over 4} \left( v^2 +
\delta v_+^2(\mu)\right) W^+_\mu {W^-}^\mu +
{1\over 2} W^3_\mu \partial^2 {W^3}^\mu + {1\over 2} B_\mu \partial^2 B^\mu
+\cr
+&{1\over 2} \left(g_3(\mu) W_3^\mu - g'(\mu)  B^\mu \right)\left[{1\over 4}
\left(v^2+\delta v^2_3(\mu)\right)-\delta Z_{3Y}(\mu) \partial^2 \right]
\left(g_3(\mu) {W_3}_\mu - g'(\mu) B_\mu \right) \cr
+& \ \bar \psi \ i
\Dslash(\ g_+(\mu)W^+,\ g_3(\mu)W_3,\ g'(\mu)B)\ \psi \ ,\cr}
\eqn\two$$
where, to the order we are working, i.e. one loop:
$$\eqalign{&g_+^2(\mu)\approx g^2 \left(1-g^2 \delta Z_+(\mu)\right)\cr
&g_3^2(\mu)\approx g^2\left(1-g^2 \delta Z_3(\mu)-g^2\delta
Z_{3Y}(\mu)\right)\cr
&g'^2(\mu)\approx g'^2\left( 1-g'^2 \delta Z_Y(\mu) -g'^2 \delta
Z_{3Y}(\mu)\right)\qquad .\cr}\eqn\two$$

Notice that below the top quark mass the initially unique coupling constant
$g$ has split into $g_+$ and $g_3$ \Ref\Peccei{R.D. Peccei and S. Peris\j
Phys. Rev. &D44 (91) 809.}.
Similarly $v_+^2(\mu)\equiv v^2 +
\delta v^2_+(\mu)$ and $v^2_3(\mu)\equiv v^2+\delta v_3^2(\mu)$ are also
different. The matching
conditions are very easily obtained since they are nothing else than the
diagrams of fig. 1 evaluated at $\m=\mt$. One obtains for instance, at one
loop,
$$\eqalign{\delta v_+^2(\m)=&{N_c\over \4pi^2} \mt^2
\left(2 \log(\m^2/\mt^2) +1
\right)\quad ,\cr
\delta v_3^2(\m)=&{N_c\over \4pi^2} \mt^2 \ 2 \log(\m^2/\mt^2)\qquad . \cr}
\eqn\four$$

Therefore this means that
$$\delta v_+^2(\mt)={N_c\over \4pi^2}\ \mt^2\ ,\qquad
\delta {v_3}^2(\mt)=0\quad .\eqn\five$$

Analogously,
$$\delta Z_{3Y}(\mt)=0\ ,\quad g_+(\mt)=g_3(\mt)=g\ \ {\rm and}\ \
g'(\mt)=g'\quad .\eqn\six$$

Equation \six\ says that the coupling constants are continuous across the
threshold. This is true as long as one keeps only the leading logarithms. In
general there are non-logarithmic pieces that modify \six\ such as, for
instance, the non-log term in the first of eqs. \four. The point is that
this term in \four\ is multiplied by $\mt^2$ (i.e. a non-decoupling effect)
and therefore contributes  (in fact dominates) for large $\mt$, whereas
the same does not happen in \six. Therefore, non-log
corrections to \six\ do not affect
the large-$\mt$ discussion that follows.

In order to obtain the effective Lagrangian at the relevant lower scales
$\m\simeq M\equiv M_W, M_Z$\foot{Hence we neglect possible terms $\sim \log
M_W/M_Z$.} one has to scale this Lagrangian down using the RGE for
each "coupling" $g_+(\m),g_3(\m),g'(\m),\delta v_+^2(\m), \delta v_3^2(\m)$
and $\delta Z_{3Y}(\m)$.
The running of
$\delta v_{+,3}^2(\m)$ is zero since
it must be proportional to a light fermion mass, which we neglect\foot{We
also neglect the contribution of the gauge bosons and the Higgs since they
do not have QCD corrections. This simplifies the analysis enormously.}.
Therefore,
$$\delta v_{+,3}^2(\mt)=\delta v_{+,3}^2(M)\qquad .\eqn\seven$$

For the other runnings, one immediately obtains ($t\equiv \log \m^2$),
$$\eqalign{&{dg_+^2\over dt}=3{g_+^4\over \4pi^2}+...\qquad ,
\qquad {dg^2_3\over dt}={10\over 3} {g^4_3\over \4pi^2}+...\cr
&{dg'^2\over dt}={50\over 9} {g'^4\over \4pi^2}+...\qquad ,\qquad {d\over
dt}\delta Z_{3Y}={1\over 6 \4pi^2}+...\cr}\eqn\eight$$

{}From diagrams such as the ones in fig. 1, but with all fermions now
contributing
since, apart from top, all others are lighter than $M_Z$. Ellipses in eq.
\eight\ stand for the contribution of the gauge bosons and the
Higgs\foot{We note again that this contribution will not have QCD
corrections to the order we
are working.}.

\Ref\Grinstein{B. Grinstein and M.-Y. Wang\j Nucl. Phys. &B377 (92) 480; see
also A. Cohen, H. Georgi and B. Grinstein\j Nucl. Phys. &B232 (84) 61.}

So far $\a=0$. The incorporation of QCD corrections proceeds in two steps.
Firstly, all couplings and masses develop a dependence on $\m$ through gluon
loops, so that for instance $\mt$ becomes $\mt(\m)$. Secondly, the matching
conditions and the RGE's \five-\eight\ get corrected by an $\a$-dependent
term. In general, the calculation of this term in the matching conditions is a
hard
two-loop calculation, and thinking in terms of an EFT does not help much to
obtain this number. It has to be calculated. For the case of eq. \five ,
this was done in ref. [\Djouadi]
with the result,
$$\eqalign{{v_+^2(M)-v^2_3(M)\over v^2_+(M)}&=
{v_+^2(\mt)-v^2_3(\mt)\over v^2_+(\mt)}=\cr
&={3\over \4pi^2}{\mt^2(\mt)\over v^2_+(\mt)}\ \left[1- {2\over
9}{\a(\mt)\over \pi}(\pi^2-9) + {\cal O}(\a^2)\right]\cr}\eqn\nine$$
where, as nicely explained in ref. [\Grinstein] , the scale in $\a(\m)$ and
$\mt(\m)$ clearly has to be $\sim \mt$ (and not $M_Z$ or $m_b$) because it is
nothing but a matching condition at $\m=\mt$. This is the $\rho$ parameter.
We shall see below that $v^2_+(\mt)=(\sqrt 2 \gf)^{-1}$, where $\gf$ is the
$\m$-decay constant. In ref. [\Grinstein] eq.
\nine\ was written in terms of the pole mass $\mt$ using the following
relation connecting $\mt(\mu)$ and $\mt$:
$$\mt=\mt(\mt)\left( 1+{4\a(\mt) \over 3\pi} + 10.95 \left({\a(\mt) \over
\pi}\right)^2+...\right)\quad ,\eqn\last$$
to ${\cal O}(\a)$.
Recently\Ref\Sirlinone{A. Sirlin,
BNL preprint, hep-ph 9403282, March 1994; see also B.H. Smith and M.B.
Voloshin, TPI-MINN-94/16-T, April 1994, and TPI-MINN-94/5-T, Jan. 1994; B.
Kniehl, DESY 94-036, March 1994.}, Sirlin has noted
that because this relation has a very large coefficient accompanying the
${\cal O} (\a^2)$ contribution, one can take eq. \nine\ as a very interesting
starting point for the analysis of the ${\cal O}(\a^2)$ corrections to eq.
\nine \Ref\Tarasov{See however the recent explicit calculation of L. Avdeev,
J. Fleischer, S. Mikhailov and O. Tarasov, Bielefeld preprint BI-TP-93/60,
June 1994.}. Within the EFT approach it comes out very naturally.

Corrections of ${\cal O}(\a)$ to eq. \six\ can be disregarded because they
are subleading in our leading-log calculation. Similarly eq. \seven\ also
remains valid even when $\a\not =0$.

The inclusion of QCD corrections also modifies the RGE's \eight\ but this
modification
has been known since the times of
RG applications to grand unified theories\Ref\Jones{See for instance D.R.T.
Jones\j Phys. Rev. &D25 (82) 581. See also the exhaustive analysis of M.E.
Machacek and M.T. Vaughn\j Nucl. Phys. &B222 (83) 83\j Nucl. Phys. &B236
(84) 221\j Nucl. Phys. &B249 (85) 70.} . All it amounts to is multiply
every quark contribution to eq. \eight\  by $\left(1+\a(\m)/\pi\right)$.
Therefore one obtains,

$$\eqalign{&{dg^2_+\over dt}={g^4_+\over \4pi^2} \left[2\ \left(1 +{\a
(t)\over \pi}\right) + 1\right] +...\cr
&{dg^2_3\over dt}={g^4_3\over \4pi^2} \left[{7\over 3}\ \left(1 +{\a
(t)\over \pi}\right) + 1\right] +...\cr
&{dg'^2\over dt}={g'^4\over \4pi^2} \left[{23\over 9}\ \left(1 +{\a
(t)\over \pi}\right) + 3\right] +...\cr
&{d\over dt} \delta Z_{3Y}={1\over 6\4pi^2} \left(1 +{\a
(t)\over \pi}\right) +...\cr}\eqn\ten$$

to be supplemented with the running of $\a(t)$,
$${d\a\over dt}=-{\beta_0\over \4pi}\ \a^2\quad ,\quad \beta_0=11-{2\over 3}
n_f\ ,\
n_f=5\ {\rm flavors}\quad .\eqn\eleven$$

Given that $g_+^2,g_3^2$ and $g'^2$ are all rather smaller than $g_s^2\equiv
4\pi \a$, a reasonable
approximation is to take into account the running of $\a$ in eqs. \ten\ while
keeping the $g_+, g_3$ and $g'$ frozen at a given value. This is tantamount
to resumming
the leading log's accompanying powers of $\a$ but not those accompanied by
powers of $g_+, g_3$ and $g'$.

With all this, one can now go about computing a typical physical quantity
like for instance $\Delta r_W$ \foot{$\Delta r_W$ is the same as the more
familiar parameter $\Delta r$ defined by Marciano and Sirlin\Ref\Marciano{W.
Marciano and A. Sirlin\j Phys. Rev. &D22 (80) 2695.} but
without the running of $e(\mu)$.}. In the EFT language this is obtained
in the following way. According to the Lagrangian \two\  the physical $W$ and
$Z$ masses are given by the equations
$$\eqalign{M_W^2=&{g^2_+(M)\over 4}\ v^2_+(M)\cr
M_Z^2=&{g_3^2(M)\over 4 c^2(M)}\ \left(v^2_3(M)+ 4 M_Z^2 \delta
Z_{3Y}(M)\right)\cr}\eqn\twelve$$
where $c^2(M)=\cos^2\theta_W(M)$ and $\tan\theta_W(M)\equiv g'(M)/g_3(M)$.

Following the EFT technique, at the scale of the W mass one should
integrate out the W boson. This gives rise to the appearance of 4-fermion
operators that mediate $\m$ decay, with strength $\gf(M)/\sqrt 2$.
The matching condition therefore becomes
$${\gf(M)\over \sqrt 2}= {g^2_+(M)\over 8 M_W^2}={1\over 2
v^2_+(M)}\ ,\eqn\thirteen$$
but since $\gf(\m)$ does not run\Ref\Fermi{W. Marciano\j Phys. Rev. &D20
(79) 274; S. Dawson, J.S. Hagelin and L. Hall\j Phys. Rev. &D23 (81) 2666;
R.D. Peccei, lectures given at the TASI-88 Summer School, Brown Univ.,
Providence, Rhode Island, June 1988.} one can see that actually
$v^2_+(M)=\sqrt2 \gf$, where $\gf$ is the Fermi constant as measured in
$\mu$ decay. Therefore
$${\gf \over \sqrt2}={g^2_+(M)\over 8 M_W^2}={e^2(M)\over 8
M^2_W}\left[{g^2_+(M)\over g^2_3(M)}\ {1\over s^2(M)}\right]\ ,\eqn\fourteen$$
where $e^2(\m)$ is the running electromagnetic coupling constant. The
quantity $\Delta r_W$ is defined as
$${\gf\over \sqrt 2}={e^2(M)\over 8 M_W^2 s^2} \ (1+\Delta r_W)\quad .$$
Consequently,
$$1+\Delta r_W = {s^2\over s^2(M)}\ {g^2_+(M)\over g_3^2(M)}\eqn\fifteen$$
where $s^2\equiv 1- M^2_W/M^2_Z$ is Sirlin's combination\Ref\Sirlin{A.
Sirlin\j Phys. Rev. &D22 (80) 971.}.

Since we are only interested in resumming $\a$ corrections we can
approximate $\Delta r_W$ in eq. \fifteen\ by

$$\Delta r_W\approx {c^2-s^2 \over s^2}\ {g^2_3(M)-g^2_+(M)\over g^2} -
{c^2\over s^2}{v^2_+(\mt)-v^2_3(\mt)\over v^2}+{4M^2_Z\over v^2}{c^2\over
s^2} \delta Z_{3Y}(M)\ .\eqn\seventeen$$

Integration of eqs. \ten\ and \eleven , with the boundary conditions
\five-\seven , yields

$$\eqalign{{g^2_+(M)\over g^2_3(M)}&\approx 1 + {g^2\over \4pi^2}\left[
-{1\over 3}
\log \left({M^2\over \mt^2}\right)+
\log\left({\a(M)\over \a(\mt)}\right)^{-{4\over
\beta_0}}\right]\cr
\delta Z_{3Y}(M)&\approx {1\over 6\4pi^2}\left[ \log\left({M^2\over
\mt^2}\right) + \log\left({\a(M)\over \a(\mt)}\right)^{-{4\over
\beta_0}}\right]\cr}\eqn\eighteen$$

so that $\Delta r_W$ is, finally,
$$\eqalign{\Delta r_W &\approx -{c^2\over s^2} {3\over \4pi^2} \mt^2(\mt) \gf
\sqrt 2 \left[ 1- {2\over 9}{\a(\mt)\over \pi}(\pi^2-9)\right]+\cr
&+{g^2\over \4pi^2}\ {1\over 2}\left({c^2\over s^2}-{1\over 3}\right)
\left[\log\left({M^2\over
\mt^2}\right)+\log\left({\a(M)\over \a(\mt)}\right)^{-{4\over
\beta_0}}\right]\cr}\eqn\nineteen$$
with $\beta_0=23/3$. In the second term of eq. \nineteen\ one has actually
resummed all orders in $\a^n \log^n$. It is here that the powerfulness of
the RG and EFT has proved to be very useful. Therefore we learn that
while the term proportional to
$\mt^2(\mt)$ comes from matching, and has therefore a well-defined scale
$\m\simeq \mt$; the term proportional to $g^2$ comes from running, instead,
which in turn means that it has to depend on the two
scales between which
it is running, $\m\simeq M$ and $\m\simeq \mt$. From the point of view of an
EFT aficionado, eq. \nineteen\ is somewhat unconventional in that it considers
matching conditions (the $\a(\mt)$ term) together with running (the
$\a(M)/\a(\mt)$ term) both at one loop. From the QCD point of view the
former is a next-to-leading-log term whereas the latter is a leading-log
one. The reason for
taking both into account is of course that the $\a(\mt)$ term is multiplied
by the  $\mt^2\gf$ combination, which is large.

If one takes the $\a(M)/\a(\mt)$ logarithmic term, expands it in powers
of $\a$ and uses eq. \last\  to rewrite eq. \nineteen\ in terms of the pole
mass, one of course reobtains eq. \A  to the given order.

\ack

This paper is the result of very many interesting discussions with Arcadi
Santamaria. I am most grateful to him for sharing his insight with me and
for continuous encouragement. I am also very grateful to Rolf Tarrach and
Thomas Mannel for
very instructive discussions. I have also benefited from interesting
conversations with Dima Bardin, Sergio Fanchiotti, Mattias Jamin and Eduard de
Rafael. Finally, I would
also like to thank Fred Jegerlehner for a useful correspondence and Maurizio
Consoli, Cliff Burgess and Jochum van der Bij for reading the manuscript and
discussions.

\endpage

\centerline{FIGURE CAPTIONS}

Fig. 1. Diagrams contributing to the matching conditions, eqs. \five-\six.

\refout

\end